\documentclass[twocolumn,floatfix,prb,preprintnumbers,amsmath,amssymb,superscriptaddress]{revtex4}
\usepackage{graphicx}
\usepackage{dcolumn}
\usepackage{bm}
\usepackage{xspace}
\usepackage[normalem]{ulem}
\usepackage{units}
\usepackage{dcolumn}
\usepackage{paralist}
\usepackage{natmove}
\usepackage{natbib}
\usepackage{multirow}
\usepackage{amsmath}
\usepackage{amssymb}
\usepackage{amsfonts}
\usepackage{listings}
\usepackage{subfigure}
\usepackage{hyperref}
\usepackage{xcolor}
\hypersetup{
    colorlinks,
    linkcolor={blue!90!black},
    citecolor={blue!90!black},
    urlcolor={blue!80!black}
  }

\newcommand{\vecr}{{\ensuremath{\mathbf{r}}}}

\newcommand{\xtp}{VOTCA-XTP\xspace}

\newcommand{\orca}{ORCA\xspace}
\newcommand{\nwchem}{NWChem\xspace}

\newcommand{\equ}[1]{Eq.~(\ref{equ:#1})}

\newcommand{\tab}[1]{Tab.~\ref{tab:#1}}

\newcommand{\fig}[1]{Fig.~\ref{fig:#1}}
\newcommand{\Fig}[1]{Figure~\ref{fig:#1}}
\newcommand{\sect}[1]{Sec.~\ref{#1}}
\newcommand{\Sect}[1]{Section~\ref{#1}}

\newcommand{\ket}[1]{{\ensuremath{\vert #1 \rangle}}}
\newcommand{\bra}[1]{{\ensuremath{\langle #1 \vert}}}

\newcommand{\ks}{\ensuremath{\text{KS}}}

\newcommand{\qp}{\ensuremath{\text{QP}}}
\newcommand{\gw}{\ensuremath{{GW}}\xspace}
\newcommand{\gwmm}{{\ensuremath{GW\!/\text{pMM}}}\xspace}

\newcommand{\alphamadn}{$\alpha$-MADN\xspace}
\newcommand{\betamadn}{$\beta$-MADN\xspace}
\newcommand{\deposit}{DEPOSIT\xspace}

\newcommand{\amadn}{$\alpha$-MADN\xspace}
\newcommand{\bmadn}{$\beta$-MADN\xspace}

\newcommand{\newtext}[1]{}

\newcommand{\SI}{Supplemental Material\xspace}

\begin{document}

\title{Quantitative Predictions of Photoelectron Spectra in Amorphous
  Molecular Solids from Multiscale Quasiparticle Embedding}

\author{Gianluca Tirimb\`{o}}
\email{g.tirimbo@tue.nl}
\affiliation{Department of Mathematics and Computer Science, Eindhoven University of Technology, P.O. Box 513, 5600 MB Eindhoven, The Netherlands}
\affiliation{Institute for Complex Molecular Systems, Eindhoven University of Technology, P.O. Box 513, 5600 MB Eindhoven, The Netherlands}

\author{Xander de Vries} 
\affiliation{Department of Applied Physics, Eindhoven University of Technology, P.O. Box 513, 5600 MB Eindhoven, The Netherlands}

\author{Christ H.L. Weijtens} 
\affiliation{Department of Applied Physics, Eindhoven University of Technology, P.O. Box 513, 5600 MB Eindhoven, The Netherlands}

\author{Peter A. Bobbert} 
\affiliation{Department of Applied Physics, Eindhoven University of Technology, P.O. Box 513, 5600 MB Eindhoven, The Netherlands}
\affiliation{Institute for Complex Molecular Systems, Eindhoven University of Technology, P.O. Box 513, 5600 MB Eindhoven, The Netherlands}

\author{Tobias Neumann} 
\affiliation{Nanomatch GmbH, Griesbachstr. 5, 76185 Karlsruhe, Germany}

\author{Reinder Coehoorn} 
\affiliation{Department of Applied Physics, Eindhoven University of Technology, P.O. Box 513, 5600 MB Eindhoven, The Netherlands}
\affiliation{Institute for Complex Molecular Systems, Eindhoven University of Technology, P.O. Box 513, 5600 MB Eindhoven, The Netherlands}

\author{Bj\"orn Baumeier} 
\thanks{Corresponding author} 
\email{b.baumeier@tue.nl} 
\affiliation{Department of Mathematics and Computer Science, Eindhoven University of Technology, P.O. Box 513, 5600 MB Eindhoven, The Netherlands}
\affiliation{Institute for Complex Molecular Systems, Eindhoven University of Technology, P.O. Box 513, 5600 MB Eindhoven, The Netherlands}

\begin{abstract}
We present a first-principles-based multiscale simulation framework for quantitative predictions of the high-energy part of the Ultraviolet Photoelectron Spectroscopy (UPS) spectra of amorphous molecular solids. The approach combines a deposition simulation, many-body Green's Function Theory, polarizable film-embedding, and multimode electron-vibrational coupling and provides a molecular-level view on the interactions and processes giving rise to spectral features. This insight helps bridging the current gap between experimental UPS and theoretical models as accurate analyses are hampered by the energetic disorder, surface-sensitivity of the measurement and the complexity of excitation processes. In particular this is relevant for the unambiguous determination the highest occupied molecular orbital energy (HOMO) of organic semiconductors, a key quantity for tailoring and engineering new opto-electronic devices. We demonstrate the capabilities of the simulation approach studying the spectrum of two isomers of 2-methyl-9,10-bis(naphthalen-2-yl)anthracene (MADN) as archetypical materials showing a clearly separated HOMO peak in experiment. The agreement with experiment is excellent, suggesting that our approach provides a route for determining the HOMO energy with an accuracy better than \unit[0.1]{eV}.
\end{abstract}

\maketitle

\section{Introduction}
\label{sec:introduction}
Amorphous organic semiconductors are intensively applied in opto-electronic devices such as organic light-emitting diodes (OLEDs)~\cite{tang_organic_1987,kido_multilayer_1995,baldo_highly_1998,adachi_third-generation_2014}, photovoltaic cells~\cite{halls_efficient_1995,yu_polymer_1995,zhao_molecular_2017} and photodetectors~\cite{peumans_small_2003,jansen-van_vuuren_organic_2016,kielar_organic_2016}. Device properties can be tuned by varying the chemical building blocks or the material processing conditions, and by combining different molecular materials in complex blends or layer stacks. A key parameter determining the functioning of a material in a device is the ionization energy, often termed the highest occupied molecular orbital (HOMO) energy, $\varepsilon_{\textrm{HOMO}}$. Relative changes of $\varepsilon_{\textrm{HOMO}}$ on the order of \unit[0.1]{eV} can already significantly alter the charge transport through host-guest materials or across internal interfaces between layers. However, even when using the perhaps most direct method for measuring $\varepsilon_{\textrm{HOMO}}$, Ultraviolet Photoelectron Spectroscopy (UPS), this level of accuracy has so far not been accomplished. 

Excitation processes in organic semiconductors are complex because their localized nature gives rise to strong structural reorganization (polaron formation) and electron-vibration coupling, which leads to shifts, broadening and additional features in the UPS spectrum~\cite{salaneck_temperature-dependent_1980,malagoli_multimode_2004,kera_first-principles_2009,kera_photoelectron_2015}. Combined with the energetic disorder originating from the amorphous structure and the surface-sensitivity of the measurement, this obstructs the unambiguous analysis of the spectra. As a result, the method used for deducing $\varepsilon_{\textrm{HOMO}}$ from the spectra (from the first peak energy or from an effective onset energy?) is a subject of debate~\cite{hill_charge-separation_2000,dandrade_relationship_2005,krause_determination_2008}. This uncertainty hampers the use of UPS studies for the rational design of new devices, and the combined use of high-resolution UPS, inverse UPS and photoluminescence spectra for obtaining accurate exciton binding energies~\cite{yoshida_complete_2015}. Qualitative understanding of the spectra is often sought via gas-phase single-molecule calculations based on density-functional theory (DFT). However, the obtained energy levels need to be artificially shifted and broadened due to the well-known underestimation of the single-particle energy gap by DFT~\cite{perdew_density_1985,mori-sanchez_localization_2008} and due to the effects of intermolecular interactions~\cite{tadayyon_reliable_2004}. Those calculations lack an explicit link to the molecular morphology, cannot resolve surface and bulk contributions to the density of states (DOS), and do not account for the spectral consequences of the molecular ionization process resulting from the excitation of molecular vibrations. This lack of predictive power combined with the ambiguity in extracting the DOS from the experimental data is a big obstacle for the development of layer stacks for organic photovoltaics or next-generation OLEDs, for which the functioning and ultimate performance is already sensitive to energy level variations of only \unit[100]{meV}.

In this work we present a first-principles-based multiscale simulation approach that bridges the current gap between experimental UPS and theoretical models by providing a quantitative prediction of the high-energy part of the UPS spectrum from which the ionization potential is derived. It consists of an accurate evaluation of (i) quasiparticle energy levels within the $GW$ approximation and (ii) thin-film embedding effects, using a hybrid quantum-mechanics/molecular-mechanics (QM/MM) approach that takes the molecular polarizabilities and the long-range interactions due to partially ordered static multipole moments into account, (iii) the inclusion of surface sensitivity via the electron attenuation length (EAL), $\Lambda$, and (iv) a full-quantum treatment of electron-vibration coupling. We focus here on only one type of initial state (frontier orbital), and simulate the UPS spectrum for perpendicular emission as a weighted sum of individual molecular environment-dependent densities of states according to
\begin{equation}
S_{\text{UPS}} (E) = \frac{1}{N_\text{m}}\sum_{j=1}^{N_\textrm{m}}
S_\text{el-vib}(E;\varepsilon_{j})
\exp\bigg({-\frac{z_0(x,y)-z_{j}}{\Lambda}}\bigg) ,
\label{equ:signal}
\end{equation}
with $\varepsilon_{j}$ the frontier orbital energy level of molecule~$j$, $z_j$ the distance of the molecule's center-of-mass (COM) to the corrugated surface at $z_0(x,y)$~\cite{note_surface}, and $S_\text{el-vib}(E;\varepsilon_{j})$ the energy-dependent spectral shape due to electron-vibration coupling. $N_{\textrm{m}}$ is the total number of molecules included in the summation, which is equal to the number of molecules for which from a vapour deposition simulation the atomistic morphology is obtained (see \sect{sec:atomistic}). In view of the large optical absorption depth, we neglect optical matrix element effects.

\begin{figure}[tb]
  \centering
  \includegraphics[width=\linewidth]{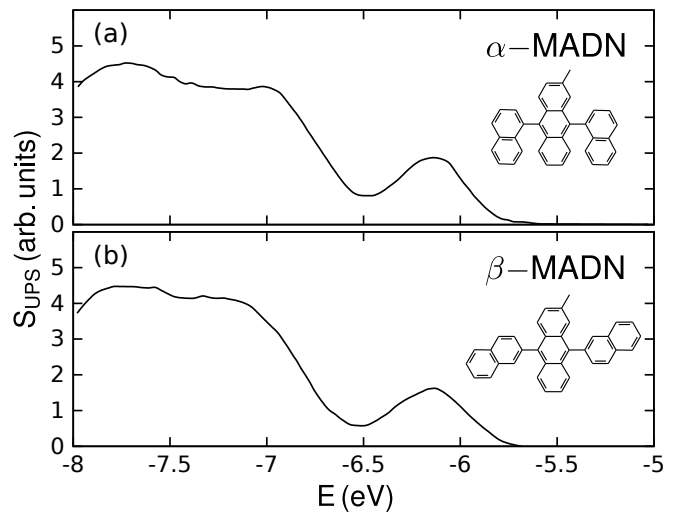}
  \caption{Experimental UPS spectra obtained using He-I radiation
    (\unit[21.2]{eV}) for (a) \alphamadn and (b) \betamadn, respectively. For both systems, the peak associated to the HOMO, at about \unit[-6.1]{eV}, is clearly separated from the deeper levels.}
\label{fig:experiment}
\end{figure}

As prototypical systems, we study the UPS spectrum for thin films of the $\alpha$ and $\beta$ isomers of 2-methyl-9,10-bis(naphthalen-2-yl)anthracene (MADN), whose chemical structures are shown as insets in~\fig{experiment}. MADN is a morphologically stable amorphous wide-gap semiconductor~\cite{lee_stable_2004} that is used extensively as an ambipolar host material in OLEDs containing deep blue fluorescent emitter molecules~\cite{lee_highly_2005,liao_highly_2005,ho_highly_2006,huang_development_2012,ho_39.2_2010,ahn_highly_2018}. The methyl substituent disrupts the symmetry and stabilizes the material against crystallization. The type of coupling of the anthracene core and the naphtyl substituents ($\alpha$ or $\beta$) affects the planarity of the molecules, and thereby the frontier orbital energies and their distribution in a thin film. We regard MADN as particularly suitable for this study because it exhibits a HOMO peak that originates from a single non-degenerate state, located predominantly on the anthracene core. Experimental high resolution (low instrumental broadening) UPS measurements (see \fig{experiment} and Sect.~S1 of the Supplemental Material~\cite{suppmat}), show that the peak full width at half maximum ($\unit[\sim 0.4]{eV}$) is significantly smaller than for many other often-used hole transporting and emitting materials in OLEDs.  It is furthermore advantageous that the HOMO state is well-separated from the deeper levels. The selection of the two isomers enables us to study the effects of morphology differences and the related effects on energy level shifts due to the small molecular dipole moments.

The paper is organized as follows: In \sect{sec:methodology}, we provide an overview of the methodology and the computational details, including the specifics of the simulation of the atomistic thin-film morphologies, the calculation of quasiparticle energies in the $GW$ approximation, the QM/MM quasiparticle embedding schemes, and the carrier-vibration coupling.  \Sect{sec:resultsanddiscussions} contains the results obtained for the thin films of $\alpha$- and $\beta$-MADN, and a discussion of the effects of long-range interactions on the quasiparticle energies, the layer-resolved and surface density-of-states, and the final UPS simulations including carrier-vibration coupling. A brief summary (\Sect{sec:summaryandconclusions}) concludes the paper.
\section{Methodology}
\label{sec:methodology}
\subsection{Atomistic thin-film morphologies}
\label{sec:atomistic}
As a first step, realistic thin-film morphologies are obtained using the Metropolis Monte Carlo-based simulated annealing protocol \deposit~\cite{neumann_modeling_2013}. It mimics the vapor deposition (PVD) process and provides molecular morphologies that exhibit commonly observed PVD characteristic features~\cite{friederich_molecular_2017,friederich_built-potentials_2018}.  Each molecule was deposited using 32 simulated annealing cycles with 120000 Monte-Carlo steps each, with annealing temperatures decreasing from \unit[4000]{K} to \unit[300]{K}. Periodic boundary conditions were applied in the directions perpendicular to the growth direction with a side length equal to \unit[10]{nm}. The deposition substrate is represented by a fixed dense layer of MADN. The energy at each simulation step was computed using customized force-fields generated using the Parametrizer module of the \deposit code.  These force-fields comprise Coulomb electrostatics based on partial charges obtained from an electrostatic potential fit~\cite{besler_atomic_1990}, Lennard-Jones potentials to account for the van-der-Waals interaction and the Pauli repulsion, and compound-specific dihedral force-fields with quantum chemistry accuracy generated by the Dihedral Parametrizer of the \deposit code. The final deposited morphologies contain 1000 molecules and are about \unit[10]{nm} thick. For the following analysis, we remove the bottom \unit[2]{nm} of the film to avoid spurious effects from the artificial substrate.

\subsection{Quasiparticle Energies in the GW Approximation}
\label{sec:quasiparticles}
Next, the internal contributions to the HOMO energy of all individual molecules are calculated including quasiparticle corrections within the $GW$ approximation of many-body Green's Functions theory~\cite{sham_many-particle_1966,hedin_effects_1970}. At this level, the calculations already include the effects of molecular deformations, obtained from the morphology simulations, but not yet the effects of embedding in the polarizable thin-film environment. These will be discussed in \Sect{sec:envirorment}. Throughout this subsection, atomic (Hartree) units are used ($\hbar = 1$, $m_{e} = 1$ and $e^2/(4\pi\epsilon_0) = 1$, with $m_e$ the electron mass, $e$ the elementary electron charge, and $\epsilon_0$ the vacuum permittivity).

The properties of a closed shell system of $N$ electrons with spin singlet ground state  can be calculated using DFT by solving the \emph{Kohn--Sham} (KS) equations~\cite{onida_electronic_2002}
\begin{equation}
  [T_{0} + V_\text{ext} + V_\text{Hartree}
  +V_\text{xc}]\ket{\phi^\ks_{i}} = \epsilon^\ks_{i} 
  \ket{\phi^\ks_{i}} ,
  \label{equ:theory:kse}
\end{equation}
with $T_0$ the kinetic energy operator, $V_\text{ext}$ an external potential, $V_\text{Hartree}$ the Hartree potential, $V_\text{xc}$ the exchange-correlation potential, and $\epsilon^\ks$ ($\ket{\phi^\ks}$) the KS energies (wave functions), respectively. Particle-like excitations, known as quasiparticles (QP), in which one electron is added to or removed from the $N$-electron ground state, are described by the one-body Green's function, $G_1$~\cite{sham_many-particle_1966,hedin_effects_1970} that obeys a Dyson-type equation of motion. This, in spectral representation, is given by
\begin{equation} 
  [ H_0 + \Sigma(E) ]\, G_1(E) = E \, G_{1} (E)\, ,
  \label{equ:theory:eqm}
\end{equation}
with $H_0 = T_{0} + V_{\text{ext}} + V_{\text{Hartree}}$ and with $\Sigma(E)$ an electron self-energy operator that describes the exchange-correlation effects. It can be shown that~\equ{theory:eqm} is part of a closed set of coupled equations, known as \emph{Hedin equations}~\cite{hedin_new_1965,strinati_application_1988}.  KS-DFT is an approximate solution for the excited-electrons problem, in which $\Sigma \sim V_\text{xc}$. A beyond-KS-DFT solution of \equ{theory:eqm} is given by the \emph{GW approximation}, in which the self energy is expressed as
\begin{equation}
  \Sigma(\vecr, \vecr', \omega) = \frac{i}{2 \pi} \int d\omega' G_{1}(\vecr,\vecr',\omega+\omega') W(\vecr,\vecr',\omega')\, .
  \label{equ:theory:pentaa}
\end{equation}
Here, $W(\vecr,\vecr',\omega')= \int \epsilon^{-1}(\vecr,\vecr'',\omega')\,v_{c}(\vecr'',\vecr')\,d\vecr''$ is the screened Coulomb interaction  between unit charges at position \vecr{} and \vecr', with $v_{c}(\vecr,\vecr') = 1 / |\vecr-\vecr'|$ being the bare Coulomb interaction and $\epsilon^{-1}(\vecr,\vecr'',\omega)$ the inverse of the frequency-dependent dielectric function. We calculate the latter in the \emph{random-phase approximation} (RPA)~\cite{hybertsen_first-principles_1985}. Employing the $GW$ approximation, we express the quasiparticle wave functions as linear combinations of KS states and obtain their energies $\varepsilon_i^\qp$ after diagonalization of the energy-dependent $GW$ Hamiltonian~\cite{aulbur_quasiparticle_2000,rohlfing_excited_2000}
\begin{equation}
H_{ij}^\gw (E) = \varepsilon^\ks_{i} \delta_{ij} +
\bra{\phi^\ks_{i}}\Sigma(E)-V_\text{xc}
\ket{\phi^\ks_{j}} .
\label{equ:theory:gwhamiltonian}
\end{equation}
In a perturbative treatment of \equ{theory:gwhamiltonian}, in which it is assumed that $\ket{\phi^\qp_i}\approx\ket{\phi^\ks_i}$, the quasiparticle energies are determined as diagonal elements of $H_{ij}^\gw$. This leads to
\begin{equation}
\begin{split}
  \varepsilon_i^{\qp,\mathrm{pert}}&= \varepsilon_i^\ks + \Delta \epsilon_i^{GW} \\
  &=
  \varepsilon_i^\ks + \bra{\phi^\ks_i}
  \Sigma(\varepsilon_i^{\qp,\mathrm{pert}})-V_\text{xc} \ket{\phi^\ks_i} \, .
\end{split}
  \label{equ:theory:gw_sc}
\end{equation}
A comparison between these two approaches is given in Sect.~S2 of the \SI~\cite{suppmat} for the two MADN derivatives.

As the self-energy is energy-dependent, and thus depends on $\varepsilon_i^\qp$, the solution to \equ{theory:gwhamiltonian} or \equ{theory:gw_sc} need to be found self-consistently. Both the correction term $\Delta \epsilon_i^{GW}$ and the non-local, energy-dependent microscopic dielectric function calculated within the RPA depend on $\varepsilon_i^\qp$~\citep{hybertsen_first-principles_1985,rohlfing_efficient_1995}. Within the \gw{} method employed in this work, we iteratively solve the Hamiltonian by updating as well the energy-dependent non-local dielectric function, until self-consistency of the eigenvalues (ev) is obtained. In literature this is often referred to as $\text{ev}GW$~\cite{wehner_electronic_2018}.

We have performed the KS-DFT and the $GW$ steps using the \orca~\cite{neese_orca_2012} and \xtp~\cite{wehner_electronic_2018} software packages, respectively. \xtp can read information from standard packages, as such as \orca, using Gaussian-type orbitals as basis functions $\left\{ \psi_i(\vecr)\right\}$ to express
\begin{equation}
 \phi^\ks_i(\vecr)=\sum_{j=0}^M X_{ij}\psi_j(\vecr).
 \label{equ:theory:basisdecomp}
\end{equation}
Matrix elements $\bra{\phi_i^\ks} V_\text{xc} \ket{\phi_j^\ks}$ needed in \equ{theory:gw_sc} are numerically integrated using spherical Lebedev and radial Euler-Maclaurin grids as used in \nwchem~\cite{valiev_nwchem_2010}, with XC functionals provided by the \emph{LibXC} library~\cite{marques_libxc_2012}.

In the evaluation of the self energy, four-center Coulomb integrals of the form
\begin{equation}
(ij|kl) = \iint d\vecr \, d\vecr' \psi_{i}(\vecr)\psi_{j}(\vecr) v_{c}(\vecr,\vecr') \psi_{k}(\vecr')\psi_{l}(\vecr')
\label{equ:theory:fourcenter}
\end{equation}
need to be calculated. \xtp makes use of the resolution-of-identity approximation (namely the RI-V approximation) to reduce the scaling from $N_\text{b}^4$ to $N_\text{b}^3$, with $N_\text{b}$ is the number of basis function. An auxiliary basis set $\{ \chi_{\nu} (\vecr) \}$ is introduced so that the integrals in \equ{theory:fourcenter}  are rewritten in the form
\begin{equation}
(ij|kl) \approx \sum_{\nu,\mu} (ij|\nu) (\nu|\mu)^{-1} (\mu|kl).
\label{equ:theory:fourcenter_RI}
\end{equation}
Here, $(\nu|\mu)^{-1}$ are elements of the inverse of the two-center Coulomb matrix $(\nu|\mu) = \iint  d\vecr_1 \, d\vecr_2 \, \chi_{\nu} (\vecr_1) v_c(\vecr_1,\vecr_2) \chi_{\mu}(\vecr_2)$ and $(ij|\nu) = \iint d\vecr_1 \, d\vecr_2 \psi_{i}(\vecr_1)\psi_{j}(\vecr_2) v_c(\vecr_1,\vecr_2) \chi_{\mu}(\vecr_2)$ are elements the three-center Coulomb matrix, respectively.

The KS-DFT eigenvalues, and thus the quasiparticle energies, may depend strongly on the exchange-correlation functional used. However, for both MADN isomers, the final $GW$ results show a negligible starting-point dependence. A comparison using the  PBE functional~\cite{perdew_generalized_1996} and the hybrid PBEh~\cite{adamo_toward_1999,ernzerhof_assessment_1999} is given in Sect.~S2 of the \SI~\cite{suppmat}.  All results reported in this paper, are obtained using the PBE functional and the cc-pVTZ basis~\cite{kendall_electron_1992} with its optimized auxiliary basis set~\cite{weigend_efficient_2002} for resolution-of-identity techniques. 

The numerical accuracy of the calculations depends on the convergence limit used, the number of levels included and the method for carrying out the frequency integration in \equ{theory:pentaa}. The convergence limit for the self-consistent $GW$-cycles in the $\text{ev}GW$ scheme was set to $10^{-5}$ Hartree (\unit[0.27]{meV}). The number of occupied and unoccupied levels taken into account for the QP calculations is 327, while for the calculation of the RPA dielectric function the full spectrum of the KS states (1385 levels) is used. The frequency integration in \equ{theory:pentaa} can be performed in \xtp using the Fully Analytical Approach~\cite{golze_gw_2019} (FAA) or a generalized plasmon-pole model (PPM)~\cite{rohlfing_efficient_1995}. The FAA expresses the frequency dependence of the self-energy in the eigenbasis of the full RPA Hamiltonian, which is in turn  evaluated in the basis of KS product states. This approach is in principle exact. However, disadvantageously, the $N_\text{b}^6$ scaling of the FAA~\cite{golze_gw_2019}  makes its application to molecules of the size of MADN computationally extremely demanding. As an alternative, the PPM allows for a fast evaluation of the self-energy. For inorganic semiconductors, the quasiparticle energy obtained using the PPM can show deviations of several tenths of \unit[]{eV} from the exact result~\cite{golze_gw_2019,larson_role_2013}. However, for geometry-optimized \alphamadn and \betamadn, we find a difference of only \unit[0.02]{eV} between the HOMO energies obtained using the FAA and PPM approaches. All things considered, the results reported in this paper have been obtained using the PPM.

\subsection{QM/MM quasiparticle embedding schemes}
\label{sec:envirorment}
Intermolecular interactions in the thin-film environment give rise to additional non-uniform modifications of the quasiparticle energies. We determine the respective corrections to the intramolecular $GW$ energies in a coupled quantum-classical (QM/MM) procedure (``quasiparticle embedding'')~\cite{may_can_2012,baumeier_electronic_2014,schwabe_pericc2_2012,li_combining_2016,li_accurate_2018, wehner_electronic_2018}. Within the MM model, we employ a classical representation of the molecular electrostatic potential based on static and induced multipole moments, located on each of the atoms in the system. A region treated on QM (here $GW$) level is coupled to suitably defined MM regions with a special scheme, which properly includes the long-range character of electrostatic interactions among the excited and neutral  MADN molecules. Below, the details of the approach are discussed.

The classical environmental contribution to the potential (the MM part) follows from the static atomic multipole moments $Q^a_t$~\cite{stone_theory_2013}, where $t$ indicates the multipole rank and $a$ the associated atom in the molecule $A$, and from the induced moments $\Delta Q_t^a$ due to the field generated by moment $t'$ of atom $a' \neq a$ in molecule $A$ and the one generated by the moment $u$ of atom $b$ in molecule $B$:
\begin{equation}
\Delta Q_{t}^{a} = - \sum_{b \, \in B}\sum_{\substack{a' \in A\\a'\neq a}}\alpha_{tt'}^{aa'}  T_{t'u}^{a'b} (Q_u^{b} + \Delta Q_u^{b}) .
\label{equ:induced}
\end{equation} 
with $\alpha_{tt'}^{aa'}$ the atomic polarizability on each site and $T_{t'u}^{a'b}$ the tensor describing the interactions between the multipoles moments $Q^{a'}_{t'}$ and $Q^b_u$. A repeated-index summation convention is used for the multipole indices $t, t', u$. The classical total energy due to the interaction between molecules with atomic indices $a$ and $b$ in regions $\mathcal{A}$ and $\mathcal{B}$ (case $\mathcal{A} =  \mathcal{B}$ included) is given for the state $(s)$ (i.e neutral ($s=\text{n}$) or charged ($s=\text{qp}$, via quasiparticle excitation) by~\cite{stone_theory_2013}
\begin{equation}
E_\text{MM}^{(s)} = \frac{1}{2}\sum_{a \in \mathcal{A}}\sum_{b \in \mathcal{B}}{(Q_t^{a(s)} + \Delta Q_t^{a(s)}) T_{tu}^{ab} Q_u^{b(s)}}. 
\label{equ:EMM}
\end{equation}
Static atomic partial charges from a CHELPG fit to the neutral molecule's electrostatic potential~\cite{breneman_determining_1990} are used for the classical representation of molecules, and atomic polarizabilities, optimized to reproduce the polarizable volume of the molecule obtained from DFT, account for polarization effects via the induction of atomic dipoles (Thole model \cite{stone_theory_2013}).

Various schemes for coupling a QM-treated inner region to a MM-treated outer region have been described in the literature. Within an {\em additive scheme}~\cite{cao_difference_2018} (here termed $GW/\text{aMM}$), the potential of the MM environment is explicitly included in the $GW$ calculation as an additional external potential to the Hamiltonian. The QM region is directly polarized by the multipole distribution (and vice versa) and coupled solutions are found self-consistently~\cite{wehner_electronic_2018}. Within a {\em subtractive scheme} ($GW/\text{sMM}$), the QM region is replaced by a MM representation of the different states and a purely classical energy correction $E_\text{MM}^{(\text{n})} - E_\text{MM}^{(\text{qp})}$ is added to the $GW$ vacuum energies. When applying this scheme, we approximate the state of the MADN molecules after the creation of an hole by that of the cation. In both cases, the MM environment includes all molecules inside a region within a cutoff distance $r_\text{c}$ around the QM molecule. 

However, both cutoff-based techniques rely on the assumption that only short-ranged local interactions affect the energies of the QM region. In the thin-films of MADN, this is not the case as the deposition simulations reveal a weak net ordering of the small molecular dipole moment (\unit[0.59]{D} and \unit[0.56]{D} for the $\alpha$ and $\beta$ isomers). For the simulated morphologies considered, the cumulative calculated electric dipole moment parallel to the surface normal ($z$-direction) of in total \unit[65.6]{D} ($\alpha$-MADN) and \unit[35.7]{D} ($\beta$-MADN). A detailed analysis of the respective distributions resolved on molecular scale is available in Sect.~S3 of the \SI~\cite{suppmat}. The two aforementioned cutoff-based approaches cannot account for long-range electrostatic effects~\cite{kratz_long-range_2016} that result as a combination of the thin-film geometry and cumulative electrostatics. Our final calculations are therefore based on a third scheme, here termed \gwmm. This is an extension of the $GW/\text{sMM}$, in which the long-range electrostatic interaction effects are included via an infinite periodic embedding based on the traditional classical Ewald summation method~\cite{arnold_efficient_2005}. 

We show in \Sect{sec:resultsanddiscussions} that this scheme allows us to include electrostatic interactions up to an arbitrarily large cutoff distance. When only the  static point charges are considered in both regions, we call this the "static \gwmm scheme". The final results are obtained by also including polarizable (polar) interactions up to a cut-off distance $r_c^p$ of \unit[3]{nm} ("polarizable \gwmm scheme"). Outside that radius, $\alpha_{tt}^{aa'}=0$. Including these polarizable interactions reveals that the molecular dipole moments are slightly screened, so that the total accumulated dipole moment in the film is reduced by \unit[3.5]{D} ($\alpha$-MADN) and \unit[2.8]{D} ($\beta$-MADN). 

\begin{figure}[tb]
  \centering
 \includegraphics[width=0.8\linewidth]{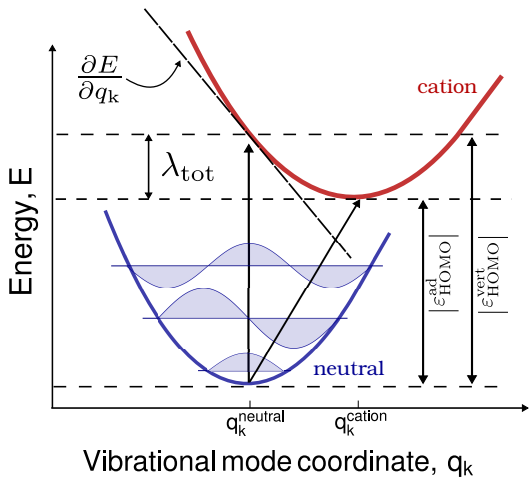}
  \caption{Schematic representation of the potential energy surfaces for the neutral (blue) and the cation (red) states, respectively, for a specific vibrational mode $k$, as a function of the (dimensionless) vibrational coordinate $q_k$. Adiabatic transitions are the results of coupling of particle-like excitations (vertical transition) with molecular vibrations, observed as the total reorganization energy $\lambda_\text{tot}= \sum_{k} \lambda_{k}$. Within the full-quantum treatment of carrier-vibrational mode coupling, employed in this paper, the detailed mode-specific coupling strengths $\lambda_k$ are included in the expression for the spectral shape shift and broadening function $S_\text{el-vib}^\text{FQ}\big(\Delta E \big)$ (\equ{fqvibrations}).}
\label{fig:vibrationalscheme}
\end{figure}

\subsection{Surface Density of States and Carrier-Vibration Coupling }
As a final step, we calculate from quasiparticle energies the HOMO contribution to the UPS spectrum, using \equ{signal}, by taking the surface-sensitivity of the experiment and the carrier-vibration coupling into account. In the absence of the carrier-vibration coupling, the spectrum would be proportional to the surface density-of-states (SDOS). This is obtained by weighting the $z$-dependent frontier orbital energy with the exponential function from~\equ{signal}, i.e.
\begin{equation}
\text{SDOS}(E) = \frac{1}{N_\text{m}} \sum_j{\delta(E-\varepsilon_j(z_j))\exp\left({-\frac{z_0(x,y)-z_{j}}{\Lambda}}\right)} .
\label{equ:sdosequation}
\end{equation}
We will adopt a value of $\Lambda=\unit[1]{nm}$ as suggested from experiment. It is compatible with estimates of the inelastic mean free path of the electrons within the random-phase approximation using $GW$ energies, which we consider an upper limit to the electron attenuation length (see Sect.~S4 of the \SI~\cite{suppmat} and Ref.~\cite{christ_2019}). The SDOS does not include the effect of the intramolecular reorganization process upon charge removal and the associated shift and lineshape broadening via carrier-vibration coupling. Conventionally, the effect of the intramolecular reorganization process upon charge removal and the associated shift and lineshape broadening via carrier-vibration coupling is described using semi-classical Marcus theory~\cite{gerischer_reaction_1976,pope_electronic_1999}. The spectral shape due to the coupling of the photoelectrons with vibrational modes is then given by
\begin{equation}
	S_\text{el-vib}^\text{Marcus} \big(\Delta E \big)=
	\frac{1}{\sqrt{4 \pi \lambda_\text{tot} k_{B} T}} \exp \left( - \frac{(\Delta E -  \lambda_\text{tot})^2}{4 \lambda_\text{tot} k_{B} T} \right),
\label{equ:marcusvibrations}
\end{equation}
with $\lambda_\text{tot}$ total reorganization energy, $k_\text{B}$ the Boltzmann constant and $T$ the temperature. The energy difference is defined relative to the adiabatic excitation energy, $\varepsilon_\text{HOMO}^\text{ad}  =  \varepsilon_\text{HOMO}^\text{vert} + \lambda_\text{tot}$ (see the schematic representation in~\fig{vibrationalscheme}). However, significant coupling with vibrational modes with energies well above $k_{\mathrm{B}} T$, such as the C-C stretch vibrations of the phenyl rings in the \unit[0.1] -- \unit[0.2]{eV} range, makes the semiclassical approach for most organic semiconductor materials invalid. This has been demonstrated for the related problems of the rates of electron or hole hopping and exciton transfer~\cite{de_vries_full_2018,DeVries2019}. Analogous to the full-quantum (FQ) approach for inter-molecular charge transfer~\cite{de_vries_full_2018}, which approximates the potential energy surface of the excited molecule in the independent mode displaced harmonic oscillator model~\cite{petrenko_analysis_2007}, we find that the spectral shape due to coupling of the photoelectrons with vibrational modes $k$ with an energy $\hbar \omega_k$ and a coupling energy $\lambda_k$ is given by
\begin{equation}
	S_\text{el-vib}^\text{FQ} \big(\Delta E \big)=
	\frac{1}{2\pi\hbar} \int_{-\infty}^{+\infty} e^{i \frac{\Delta E}{\hbar} t} e^{-F(0)} e^{F(t)} \,dt ,
\label{equ:fqvibrations}
\end{equation}
where $F(t) = \sum_{k} \frac{\lambda_{k}}{\hbar \omega_k} \left(\coth\left(\frac{\hbar \omega_k}{2 k_{\text{B}}T} \right)\cos(\omega_{k}t) + i \sin(\omega_{k}t) \right)$. Evaluating the needed parameters (vibrational modes $\omega_{k}$ and coupling energy $\lambda_k$) can be simplified under the assumptions that (i) the ground- and excited-state potential energy surfaces are harmonic and (ii) no vibrational frequency alteration or normal mode rotation occurs in the excited state. With these two assumptions, equivalent to the premise of a linear electron-phonon coupling, only the ground-state vibrational modes frequencies $\omega_{k}$ and the gradient of the total energy for the charged (excited) system in the ground state geometry with respect to the phonon mode coordinates $q_k$, $\frac{\partial E}{\partial q_k}$,  need to be evaluated as indicated in~\fig{vibrationalscheme}. The mode-specific coupling energies are then determined as $\lambda_k = \frac{1}{2} \frac{\partial E}{\partial q_k}$.

\section{Results and discussion}
\label{sec:resultsanddiscussions}

\begin{figure}[tb]
  \centering
  \includegraphics[width=\linewidth]{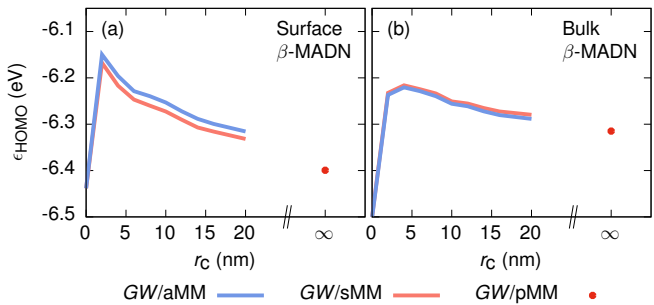}
	\caption{Comparison of the quasiparticle HOMO energies $\varepsilon_{\text{HOMO}}$ obtained using different molecular mechanics embedding schemes, as a function of the cutoff radius $r_c$ up to which electrostatic effects are included, for (a) a surface and (b) a bulk molecule in the $\beta$-MADN thin film. For the cutoff-based $GW$/aMM (blue line) and $GW$/sMM (red line) methods, the cutoff length $r_\text{c}$ is varied showing only a slow convergence. The value at $r_\text{c} = \infty$ indicates the result after periodic embedding ($GW$/pMM scheme). In all cases, only the quasiparticle state is considered polarizable while the molecules in the MM region are described by static point charges.}
  \label{fig:suppinfo:embeddingcomparison}
\end{figure}

\subsection{Quasiparticle embedding and long-range interactions}
To assess the influence of long-range electrostatic interactions on the quasiparticle energies of the two MADN thin films, we analyze the results obtained with the different embedding schemes introduced in \sect{sec:envirorment}.
First, we consider the effects of the dependence of the cutoff radius in the additive $GW/\text{aMM}$ scheme. All results given in this subsection are obtained using only a static embedding approach, within which the polarizability of the molecules is switched off. We show in~\fig{suppinfo:embeddingcomparison} (blue curve) the calculated $\epsilon_\text{HOMO}$ as a function of $r_\text{c}$ for a molecule at the surface and in the bulk-like region of the $\beta$-MADN film, respectively. The slow decrease of the energy with size of the embedding region indicates, in particular for the surface molecule, that even at a cutoff of \unit[20]{nm} no converged result is obtained. Repeating the same analysis for the computationally less demanding subtractive $GW$/sMM scheme, in which we now allow the classical substitute of the QM molecule to be polarizable as it automatically is in $GW$/aMM, yields a cutoff dependence given by the red lines in~\fig{suppinfo:embeddingcomparison}. Comparison to the $GW/\text{aMM}$ data reveals deviations smaller than \unit[0.02]{eV}. Based on this good agreement, we then use this parametrization and embed the classically represented QM molecule in a periodically repeated background in the $GW$/pMM setup. The results for both the surface and the bulk molecule are shown as data points for $r_\text{c} =\infty$ in~ \fig{suppinfo:embeddingcomparison}. For the surface molecule, the periodically embedded HOMO energy is \unit[0.07]{eV} lower than as obtained from the $GW$/sMM calculation with the largest cutoff considered. In the bulk, the difference is slightly smaller, viz. \unit[0.03]{eV}. We adopt the \gwmm scheme for the following analysis for thin-film energy levels, the SDOS and the UPS spectrum. 

\begin{figure}[tb]
  \centering
  \includegraphics[width=\linewidth]{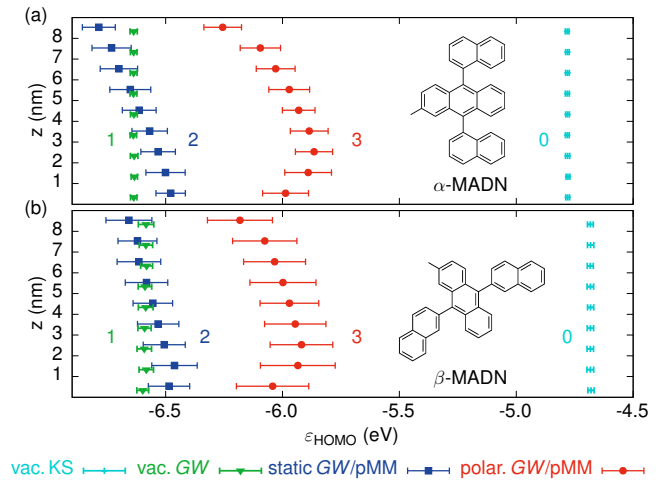}
  \caption{Layer-resolved energy levels of (a) $\alpha$-MADN and (b)
    $\beta$-MADN obtained from vacuum KS (0), vacuum
    $GW$ (1), static (2) and polarizable
    (3) \gwmm calculations, respectively. The error bars correspond to the range of $\pm$ one standard deviation.}
  \label{fig:layerprofiles}
\end{figure}

\subsection{Layer-resolved energy levels and DOS}
\label{sec:layersdos}
\Fig{layerprofiles} shows the laterally-averaged depth dependence of the HOMO energies as resulting from various levels of refinement, labeled "0" to "3". It shows that \gw corrections ("level 1") to the vacuum KS levels ("level 0") lower the energies by up to \unit[1.9]{ eV}, nearly uniformly for both isomers. The gas-phase simulations include the molecular deformations in the thin-film morphology, but these cause only a small broadening of the DOS. It is mainly due to disorder in the anthracene-naphthalene torsion angle, which is largest for $\beta$-MADN. When long-range electrostatic interactions are included (static \gwmm, level ``2", blue squares in~\fig{layerprofiles}), we find for both isomers a nearly linear $z$-dependence of the mean HOMO energy, which is symmetric with respect to the mean vacuum \gw energy $\varepsilon^\gw$. This is due to accumulating net dipole moment contributions parallel to the surface normal during film growth. Even though the dipole moments of the individual molecules are small (\unit[0.59]{D} and \unit[0.56]{D} for the $\alpha$ and $\beta$ isomers) and their average net components parallel to the growth axis are only \unit[0.065]{D} and \unit[0.040]{D}, respectively, the resulting energy gradients are a few hundredths of an \unit[]{eV/nm}. Adding polarization effects (polar \gwmm, level ``3", red circles in \fig{layerprofiles}) leads to a shift of the mean of the distributions to lower binding energies. The effect is stronger in the bulk-like center of the film (\unit[0.7]{eV} (\unit[0.6]{eV}) for \amadn (\bmadn)) than at the vacuum surface (\unit[0.5]{eV}).  The $z$-coordinate dependence of the distribution of the individual HOMO energies, obtained using the static and polarizable \gwmm schemes is shown in Sect.~S3 of the \SI~\cite{suppmat}. 

Differences between the surface and bulk energy level structure of organic materials are well-known from UPS studies. For crystalline anthracene, e.g., the experimental binding energy difference for the first and second monolayer was found to be $\unit[0.3\pm0.15]{eV}$~\cite{salaneck_intermolecular_1978}.  A similar effect is seen in~\fig{layerprofiles} for the first molecular layer near the vacuum surface, which after subtracting the energy gradient due to dipole orientation shows an increase of the binding energy of about \unit[0.15]{eV}. Since we simulate freestanding thin films, we also note modifications from the bulk-like behavior at the bottom surface with $z=0$. In view of our interest in analyzing the UPS spectrum after irradiation from the positive $z$ direction, which is only sensitive to the energy level structure in a thin zone near the top vacuum surface, we focus on that region.  

\begin{figure}[tb]
  \centering
  \includegraphics[width=\linewidth]{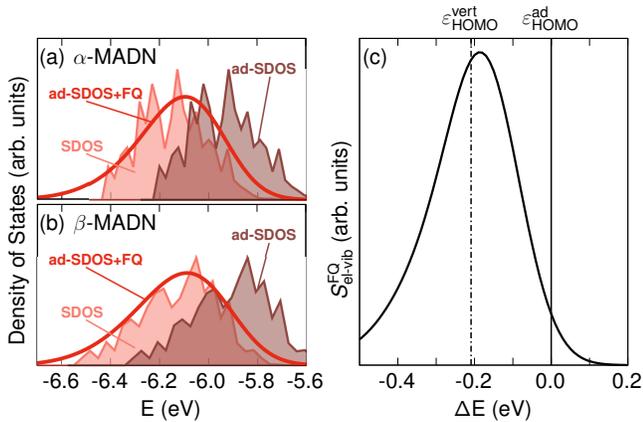}
  \caption{Frontier orbital surface density-of-states
    before (SDOS) and after adiabatic correction (ad-SDOS), as well as the simulated UPS spectra within the full-quantum model (ad-SDOS+FQ) for both \alphamadn (a) and \betamadn (b) as obtained from the polarizable \gwmm calculations. Panel (c) shows for $\beta$-MADN the shift of single-molecule vertical HOMO level to the adiabatic one at lower binding energies, and the subsequent application of the lineshape function $S_\text{el-vib}^\text{FQ}(\Delta E)$ (\equ{fqvibrations}) which leads to a pronounced broadening (FWHM: \unit[0.25]{eV}) and shift to higher binding energies with respect to $\varepsilon_\text{HOMO}^\text{ad}$.}
    \label{fig:DOSvsSDOS}
\end{figure}

\subsection{Vertical and Adiabatic Surface Density of States}
\fig{DOSvsSDOS}(a,b) show the SDOS for the two isomers (light-red shaded), obtained using~\equ{sdosequation}. The SDOS is based on the vertical excitation energies. These include the effect of the electronic polarization of the environment but do not include the effect of the intramolecular structural reorganization process upon charge removal. The ad-SDOS curves in \Fig{DOSvsSDOS}~(a,b) (dark-red shaded) show the adiabatic SDOS, obtained from the SDOS by adding an energy shift equal to the total reorganization energy, which is \unit[0.21]{eV} for both isomers. The actual excitation process is not adiabatic, but is accompanied by the excitation of vibrational modes. \Fig{DOSvsSDOS}(c) shows, for a single $\beta$-MADN molecule, the resulting shift of the DOS to a more negative HOMO energy and the resulting broadening to a full-width at half maximum of approximately \unit[0.25]{eV}. For $\alpha$-MADN, the effect is very similar. The absence of multiple satellite peaks, as usually seen for gas-phase spectra of structurally more simple molecules such as pentacene, is due to coupling of multiple modes, which smears out the structure. The peak of the function $S_{\textrm{el-vib}}^{\textrm{FQ}}$ is almost equal to the vertical HOMO energy. Including vibrational effects thus induces for these systems almost no peak shift with respect to the vertical (polarizable $GW$/pMM) energies, but only a spectral broadening. The resulting calculated UPS spectral intensity function, given by~\equ{signal}, is shown in \Fig{DOSvsSDOS}(a) and \Fig{DOSvsSDOS}(b) by the curves labeled ``ad-SDOS+FQ''.

It is interesting to compare the full-quantum vibrational response function $S_{\textrm{el-vib}}^{\textrm{FQ}}$ to the classical Marcus function. According to~\equ{marcusvibrations}, the peak energy is shifted to lower energies relative to the adiabatic excitation energy by the total reorganization energy of \unit[0.21]{eV} and the function has a Gaussian shape with a full-width at half maximum of $4\sqrt{\ln(2) \lambda_{\textrm{tot}}k_\textrm{B} T }\cong \unit[0.24]{eV}$. A comparison with \fig{DOSvsSDOS}(c) shows that in this case the full-quantum result differs only weakly from the classical Marcus response function. Apparently, the coupling of the photo-electrons to high-energy vibrational modes is for these systems relatively weak. 

\begin{figure}[tb]
  \centering
  \includegraphics[width=\linewidth]{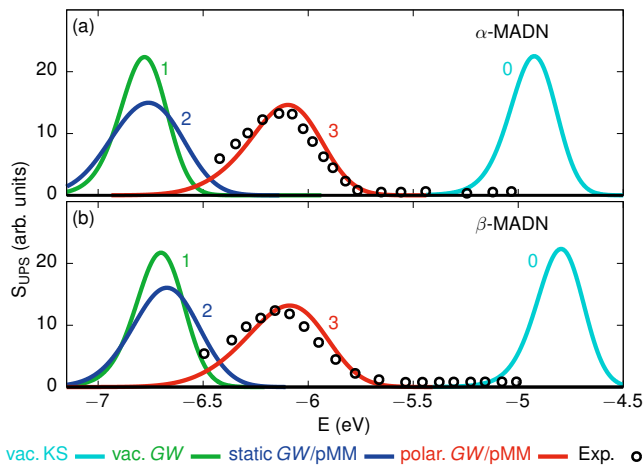}
  \caption{UPS spectra for $\alpha$-MADN (a) and $\beta$-MADN (b). Curves 0$-$3 give the UPS as depth weighted DOS with vrtical-to-adiabatic shift and vibrational broadening via \equ{fqvibrations} predicted from vacuum KS (0), $GW$ (1), static (2) and polarizable \gwmm (3) calculations, respectively. The closed circles give the experimental spectra, obtained using He-I radiation (\unit[21.2]{eV}). The experimental resolution is $\sigma$ = \unit[0.05]{eV}, and has no significant effect on the final spectral width.}
  \label{fig:DOS2UPS}
\end{figure}

\subsection{Simulated UPS including Carrier-Vibration Coupling}
\Fig{DOS2UPS} shows the final UPS spectra of the frontier orbital of $\alpha$- and $\beta$-MADN thin films, simulated for the four different levels of theory, together with the experimental data. Characteristics of the signals (maximum position, onset, and FWHM) are listed in~\tab{MADNdata}. Following the conventional approach, the onset energy is defined by extrapolating the tangent through the low-binding-energy inflection point of the HOMO peak to zero intensity. The energetic position of the simulated peaks for the different methods reflects the variations discussed for the layer-averages in~\fig{layerprofiles}. Comparison to the reference experimental spectrum now allows assessment of the quality of the various methods and the importance of the individual processes for the analysis of the experiment. 

Simulations based on vacuum energies which exclude the effects of inhomogeneous local electric fields and environment polarization either over- (KS) or underestimate ($GW$) the energy of the peak maximum by up to \unit[1.3]{eV}. The FWHM is nearly exclusively determined by the single-molecule spectral function and results about a third smaller than measured. Inclusion of static local field effects in \gwmm does not noticeably affect the peak maximum but the additional disorder contributes to further broaden the signal. Accounting for the polarization response of the material upon quasiparticle excitation in \gwmm we obtain a simulated UPS signal in excellent agreement with the measurement: the largest deviation is found for the peak maximum of $\beta$-MADN and amounts to only \unit[50]{meV}, which is also the experimental resolution. Most importantly, the comparison emphasizes that it is possible to achieve a predictive modeling of frontier orbital energies at the accuracy needed for an accurate understanding and prediction of device performance. The fact that we achieve the same accuracy by studying two isomers, with different molecular structures and thin film morphologies, supports the robustness of our approach. 

In device simulations, the bulk adiabatic ionization energy, $\varepsilon_{\textrm{HOMO,bulk}}^\text{ad}$, is needed. This energy may be obtained from a linear extrapolation of the bulk polarizable \gwmm energies shown in~\fig{layerprofiles} to the surface plane at $z = z_0$, plus the reorganization energy $\lambda_\text{tot}$ of \unit[0.21]{eV}. The resulting values, $\varepsilon_{\textrm{HOMO,bulk}}^\text{ad} = \unit[-5.91(-5.89)\pm0.05]{eV}$ for $\alpha(\beta)$-MADN, are located in between the UPS peak and onset energies (see~\tab{MADNdata}). Using either the peak or the onset value would, in this case, thus introduce an error of about \unit[0.1]{eV} or more. Our refined protocol for the analysis of UPS measurements provides a methodology for avoiding such an error. One may see from \fig{DOSvsSDOS}(c) that when carrying out a measurement for a single molecule, the onset energy would provide an excellent approximation to $\varepsilon_{\textrm{HOMO,bulk}}^{\textrm{ad}}$. However, that coincidence is fortuitous, as the peak shape and onset energies depend on the mode-resolved reorganization energies and the temperature. That may already be seen when considering the peak shape obtained within the semiclassical Marcus-theory (\equ{marcusvibrations}). The difference between the onset energy and the adiabatic ionization energy is then equal to $(-\lambda_{\textrm{tot}}+ \sqrt{8 \lambda_{\textrm{tot}} k_{\textrm{B}} T})$. For a system with $\lambda_{\textrm{tot}} = \unit[0.2]{eV}$ (close to the value for MADN) and for $k_{\textrm{B}}T=\unit[0.025]{eV}$ (close to room temperature), the onset and adiabatic ionization energies then indeed coincide. However, the electron-vibrational mode coupling shows a significant dependence on the molecules considered. The total reorganization energies vary from less than \unit[0.1]{eV} to more than \unit[0.3]{eV}, with a tendency to decrease with increasing molecular size~\cite{Devos1998,Deng2004,kera_photoelectron_2015}. Furthermore, in thin films the spectral broadening due to energetic disorder leads to a shift of the onset value to a smaller binding energy, whereas the reduced screening at the thin film surface leads to shift to a larger binding energy. For the two isomers of MADN, the former effect is largest, so that the absolute value of the onset energy is slightly smaller than $|\varepsilon_{\textrm{HOMO,bulk}}^{\textrm{ad}}|$.

\begin{table}[tb]
\centering
\caption{Characteristics of the predicted UPS spectrum (see the caption of~\fig{DOS2UPS}) for \amadn and
  \bmadn at the four different levels (0 $-$ 3) of the multiscale quasiparticle embedding
  procedure. The HOMO peak position, onset, and the full width at
  half maximum (FWHM) (all in eV) are compared to the respective
  experimental results. The table also gives the calculated bulk adiabatic ionization energy $\varepsilon_{\textrm{HOMO,bulk}}^{\textrm{ad}}$.}
\begin{ruledtabular}
\begin{tabular}{ l c c c c c }
	&\multicolumn{2}{c}{vacuum} &\multicolumn{2}{c}{\gwmm} \\
 & KS &  $GW$ & Static  & Polar. & Exp.\\
 \hline
  {$\boldsymbol{\alpha}${\bf -MADN}}   &  &  &  &  &   \\
 \hline
  $\varepsilon_{\textrm{HOMO,bulk}}^{\textrm{ad}}$  &  &  &  & -5.91 &   \\
  UPS peak   & -4.92 & -6.78 & -6.76 & -6.10 & -6.12  \\
  UPS onset  & -4.71 & -6.57 & -6.44 & -5.79 & -5.79 \\
  UPS FWHM   &  0.25 &  0.25 &  0.39 &  0.39 &  0.37 \\
 \hline
  {$\boldsymbol{\beta}${\bf -MADN}}   &  &  &  &  &   \\
  \hline
  $\varepsilon_{\textrm{HOMO,bulk}}^{\textrm{ad}}$  &  &  &  & -5.89 &   \\
  UPS peak   & -4.80 & -6.70 & -6.67 & -6.09 & -6.14\\
  UPS onset  & -4.58 & -6.48 & -6.37 & -5.75 & -5.73 \\
  UPS FWHM   &  0.25 &  0.26 &  0.36 &  0.41 &  0.42\\
\end{tabular}
\end{ruledtabular}
\label{tab:MADNdata}
\end{table}

\section{Summary and Conclusions}
\label{sec:summaryandconclusions}
In summary, we have developed a multiscale approach that provides a prediction of the frontier orbital UPS spectrum of amorphous molecular thin films. The approach includes (i) first-principles calculations of a realistic thin-film morphology, (ii) the electronic properties at a state-of-the-art (many-body Green's functions) level of quantum chemistry, (iii) embedding in a polarizable molecular mechanics environment, and (iv) the effects of the vibrational modes that are excited in the experiment. We have focused on two isomers of MADN, for which the non-degeneracy of the HOMO state leads to an exceptionally narrow width of the HOMO orbital UPS spectrum. Our work shows how the spectrum is related to the disorder-induced energy level distribution in the bulk of the organic semiconductor and near the surface. The good agreement between the calculated and experimental peak positions and widths of the UPS spectra indicates that our approach provides a route towards accurately predicting the bulk adiabatic ionization energy $\varepsilon_{\textrm{HOMO,bulk}}^{\textrm{ad}}$, which is the HOMO energy needed in device simulations.

We find that neither the HOMO energy that would follow from the UPS peak energy nor the onset energy coincide with $\varepsilon_{\textrm{HOMO,bulk}}^{\textrm{ad}}$. For the MADN films studied, both assumptions would introduce an error of about \unit[0.1]{eV} or more. Instead, the actual value of $\varepsilon_{\textrm{HOMO,bulk}}^{\textrm{ad}}$ is in this case intermediate between the onset and peak values. The error made by taking $\varepsilon_{\textrm{HOMO,bulk}}^{\textrm{ad}}$ equal to the UPS onset energy is determined by the balance between the effects of energetic disorder and reduced screening at the thin film surface. For device applications, the {\em relative} error between different materials is most important. From our study, we expect that the relative error when taking the onset energy is largest for the case of two materials with strongly dissimilar energetic disorder energies or strongly dissimilar molecular polarizabilities.

\section*{Acknowledgments}
This work has been supported by the Innovational Research Incentives Scheme Vidi of the Netherlands Organisation for Scientific Research (NWO) with project number 723.016.002 and by Horizon-2020 EU project MOSTOPHOS (Project No. 646259, XdV)


\begin{thebibliography}{73}
\bibitem [{not()}]{note_surface}%
  \BibitemOpen
  \href@noop {} {}\bibinfo {note} {The surface profile $z_{0}(x,y)$ is
  determined from a fit to the hightest COM coordinates. For surface
  corrugations $\delta z \ll \lambda_\text{in}$,
  $\exp{[-(z_0(x,y)-z_j)/\lambda_\text{in}]} \simeq
  \exp{[-(z_0-z_j)/\lambda_\text{in}]}$ in~\equ{signal}.}\BibitemShut {Stop}%
\end{thebibliography}
\end{document}


\title{Supplemental Material for ''Quantitative Predictions of Photoelectron Spectra in Amorphous Molecular Solids from Multiscale Quasiparticle Embedding''}

\author{Gianluca Tirimb\`{o}}
\email{g.tirimbo@tue.nl}
\affiliation{Department of Mathematics and Computer Science, Eindhoven University of Technology, P.O. Box 513, 5600 MB Eindhoven, The Netherlands}
\affiliation{Institute for Complex Molecular Systems, Eindhoven University of Technology, P.O. Box 513, 5600 MB Eindhoven, The Netherlands}

\author{Xander de Vries} 
\affiliation{Department of Applied Physics, Eindhoven University of Technology, P.O. Box 513, 5600 MB Eindhoven, The Netherlands}

\author{Christ H.L. Weijtens} 
\affiliation{Department of Applied Physics, Eindhoven University of Technology, P.O. Box 513, 5600 MB Eindhoven, The Netherlands}

\author{Peter A. Bobbert} 
\affiliation{Department of Applied Physics, Eindhoven University of Technology, P.O. Box 513, 5600 MB Eindhoven, The Netherlands}
\affiliation{Institute for Complex Molecular Systems, Eindhoven University of Technology, P.O. Box 513, 5600 MB Eindhoven, The Netherlands}

\author{Tobias Neumann} 
\affiliation{Nanomatch GmbH, Griesbachstr. 5, 76185 Karlsruhe, Germany}

\author{Reinder Coehoorn} 
\affiliation{Department of Applied Physics, Eindhoven University of Technology, P.O. Box 513, 5600 MB Eindhoven, The Netherlands}
\affiliation{Institute for Complex Molecular Systems, Eindhoven University of Technology, P.O. Box 513, 5600 MB Eindhoven, The Netherlands}

\author{Bj\"orn Baumeier} 
\thanks{Corresponding author} 
\email{b.baumeier@tue.nl} 
\affiliation{Department of Mathematics and Computer Science, Eindhoven University of Technology, P.O. Box 513, 5600 MB Eindhoven, The Netherlands}
\affiliation{Institute for Complex Molecular Systems, Eindhoven University of Technology, P.O. Box 513, 5600 MB Eindhoven, The Netherlands}

\maketitle
\tableofcontents
\section{Experimental details}
\label{sec:exp}
The experiments were done in a multi-chamber VG EscaLab II system with a base pressure of the deposition and the analyzer chamber in the upper $\unit[10^{-6}]{}$ and the lower $\unit[10^{-8}]{Pa}$ range respectively. $\alpha$- and $\beta$-MADN (Lumtec) were deposited by high-vacuum ($\unit[8\cdot10^{-6}]{Pa}$) thermal evaporation onto in situ sputter-cleaned Au-coated Si-substrates at a rate of about \unit[1]{nm/min}. The deposited films were transferred under ultra high vacuum (UHV) between the deposition and the analyzer chamber. The UPS spectra were recorded at a \unit[-6]{V} bias voltage using HeI radiation, generated in a differentially-pumped discharge lamp.

\section{Computational details of the GW calculations}
\label{sec:gw}
In~\Tab{supp_info:gas_phase_data} we show a comparison of the calculated gas-phase energy levels for $\alpha$-MADN and $\beta$-MADN, obtained using the PBE and PBEh exchange-correlation functionals with the cc-pVTZ basis (see the discussion and references in section II.B of the main text). In each case, the table gives the Kohn-Sham (KS) energy $\varepsilon^{\textrm{KS}}$, the perturbatively calculated $GW$ energy $\varepsilon^{\textrm{QP,pert}}$, and the exact $GW$ energy, $\varepsilon^{\textrm{QP}}$.

\begin{figure}[tb]
  \centering
  \includegraphics[width=0.75\linewidth]{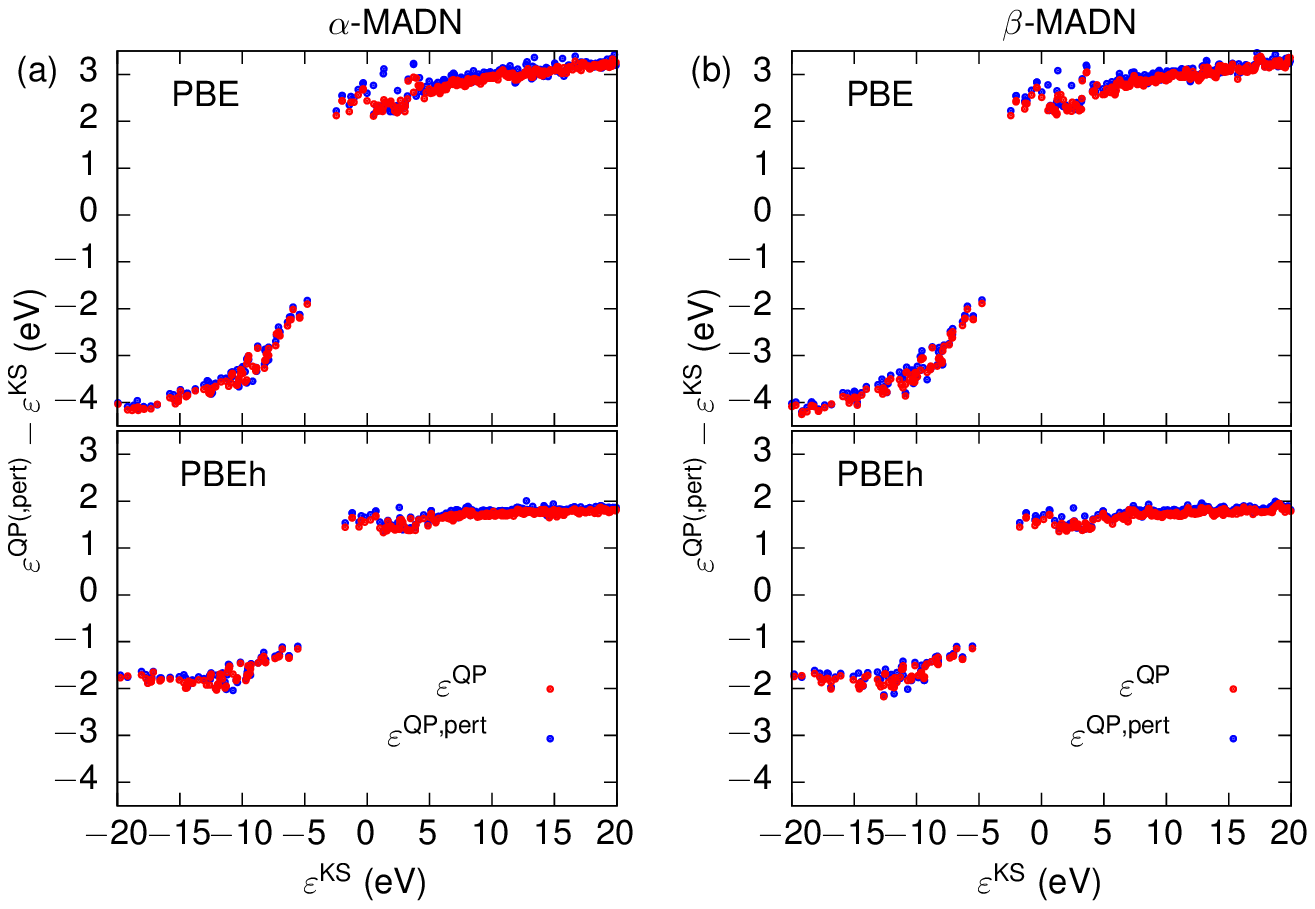}
  \caption{Calculated quasiparticle corrections as a function of the Kohn-Sham energy for (a) \alphamadn and (b) \betamadn, obtained using the PBE and PBEh DFT functionals.
  The blue and red symbols give the perturbatively calculated quasiparticle corrections ($\varepsilon^{\textrm{QP,pert}}$), and the corrections obtained after diagonalization of the quasiparticle Hamiltonian ($\varepsilon^{\textrm{QP}}$), respectively.}
\label{fig:suppinfo:selfenergy}
\end{figure}

Note the difference in the results obtained for $\varepsilon^{\textrm{KS}}$, e.g., in the HOMO energy of $\alpha$-MADN: \unit[-4.79]{eV} (PBE) and \unit[-5.55]{eV} (PBEh). This is a consequence of the spurious self-interaction in the functionals and the inadequacy of DFT to describe electronically excited states.
\Fig{suppinfo:selfenergy} shows the errors made by using the two different functionals to calculate the KS energies, as judged by comparing these energies with the perturbatively calculated or exact $GW$ energies. The correction to the KS-levels is not constant, and is quite different for the occupied and unoccupied states. We can see that for the PBE functional the quasi-particle correction is more pronounced than for hybrid PBEh functional, which already contains part of the exchange contributions to the self-energy operator. In spite of the different energy-dependent corrections, the final quasi-particle energies (see \Tab{supp_info:gas_phase_data}) do not show a significant dependence on the exchange-correlation functional used. The exact $GW$ HOMO energies, as obtained from both functionals, differ for $\alpha$-MADN ($\beta$-MADN) only by about \unit[0.01]{eV} (\unit[0.02]{eV}).

\begin{table}[tb]
\caption{Comparison of the calculated gas-phase energy levels (in \unit[]{eV}) for $\alpha$-MADN and $\beta$-MADN, obtained using the PBE and PBEh exchange-correlation functionals with the cc-pVTZ basis set, within KS-DFT ($\varepsilon^{\textrm{KS}}$), including perturbative quasiparticle corrections ($\varepsilon^{\textrm{QP,pert}}$), and after diagonalization of the quasiparticle Hamiltonian ($\varepsilon^{\textrm{QP}}$), respectively. All energies are given in units eV.}
\vspace{0.3cm}
\begin{minipage}[t]{.49\linewidth}
		\raggedleft
\begin{tabular}{l c c c }
\multicolumn{4}{c}{$\boldsymbol{\alpha}${\bf -MADN}} \\ \hline \hline
\noalign{\vskip 1mm}  
&$\varepsilon^{\textrm{KS}}$ &$\varepsilon^{\textrm{QP,pert}}$ &$\varepsilon^{\textrm{QP}}$ \\
\hline
\\
	\multicolumn{4}{c}{PBE} \\
	HOMO-2&-5.412&	-7.538&	-7.597\\
	HOMO-1&-5.396&	-7.524&	-7.595\\
	\textbf{HOMO}&\textbf{-4.786}&	\textbf{-6.610}&	\textbf{-6.689}\\
	LUMO&-2.468&	-0.245&	-0.348\\
	LUMO+1&-2.011&	0.536&	0.424\\
	LUMO+2&-2.000&	0.550&	0.435\\
\\
\multicolumn{4}{c}{PBEh} \\
	HOMO-2&-6.270&	-7.570&	-7.611\\
	HOMO-1&-6.256&	-7.559&	-7.606\\
	\textbf{HOMO}&\textbf{-5.548}&	\textbf{-6.645}&	\textbf{-6.702}\\
	LUMO&-1.755&	-0.218&	-0.313\\
	LUMO+1&-1.200&	0.544&	0.444\\
	LUMO+2&-1.189&	0.561&	0.457\\\hline
\hline
\end{tabular}
\end{minipage}
\hfill
	\begin{minipage}[t]{.49\linewidth}
	\raggedright
\begin{tabular}{l c c c }
\multicolumn{4}{c}{$\boldsymbol{\beta}${\bf -MADN}} \\ \hline \hline
\noalign{\vskip 1mm}  
&$\varepsilon^{\textrm{KS}}$ &$\varepsilon^{\textrm{QP,pert}}$ &$\varepsilon^{\textrm{QP}}$ \\
\hline
\\
	\multicolumn{4}{c}{PBE}\\
	HOMO-2&-5.459&	-7.616&	-7.690\\
	HOMO-1&-5.456&	-7.614&	-7.682\\
	\textbf{HOMO}&\textbf{-4.762}&	\textbf{-6.574}&	\textbf{-6.650}\\
	LUMO&-2.457&	-0.231&	-0.335\\
	LUMO+1&-2.030&	0.514&	0.384\\
	LUMO+2&-2.027&	0.520&	0.414\\
\\
\multicolumn{4}{c}{PBEh}\\
	HOMO-2&-6.324&	-7.652&	-7.701\\
	HOMO-1&-6.321&	-7.652&	-7.693\\
	\textbf{HOMO}&\textbf{-5.521}&	\textbf{-6.615}&	\textbf{-6.670}\\
	LUMO&-1.744&	-0.204&	-0.297\\
	LUMO+1&-1.230&	0.514&	0.411\\
	LUMO+2&-1.219&	0.533&	0.438\\
\hline
\hline
\end{tabular}\label{tab:supp_info:gas_phase_data}
\end{minipage}
\end{table}

\vspace{1cm}

\section{Quasiparticle calculations including Molecular Mechanics embedding}
\label{sec:gwmm}

\subsection{Electrostatic dipole moment distributions}
\label{sec:dipoles}
\Fig{suppinfo:dipoles}(a) and (e) show the distributions of the absolute dipole moment for $\alpha$- and $\beta$-MADN thin films, respectively, calculated based on the classical atomic point charge distributions of the constituent molecules. Both distributions are narrow and centered around the single-molecule values of \unit[0.59]{D} and \unit[0.56]{D}. From the distribution of the dipole moment's $z$-component (i.e., its component parallel to the surface normal) in~\fig{suppinfo:dipoles}(b) and (f) one can clearly see that there are more molecular dipoles aligned in positive $z$-direction. This gives rise to the observed net overall dipole moment of the thin film. Due to interactions among the molecular dipole moments as obtained from single-molecule data, it can be assumed that a more realistic description of the thin films' electrostatic properties should include mutual polarization effects. In our framework, we therefore first treat the total polarization of the film within the pMM approach described in Section II.C of the main text. After application of this background pMM, the induced moments primarily lead to a broadening of the distribution of the absolute molecular dipole moments, as can be seen in~\fig{suppinfo:dipoles}(c) and (g). Similar observations can be made for the respective $z$-component distributions in~\fig{suppinfo:dipoles}(d) and (h). The partial screening due to the polarizability of the neighbouring molecules reduces the total accumulated dipole moment in the film by \unit[3.5]{D} ($\alpha$-MADN) and \unit[2.8]{D} ($\beta$-MADN).

\subsection{Excitation energy distributions}
\begin{figure}[tb]
    \centering
    \includegraphics[width=0.49\linewidth]{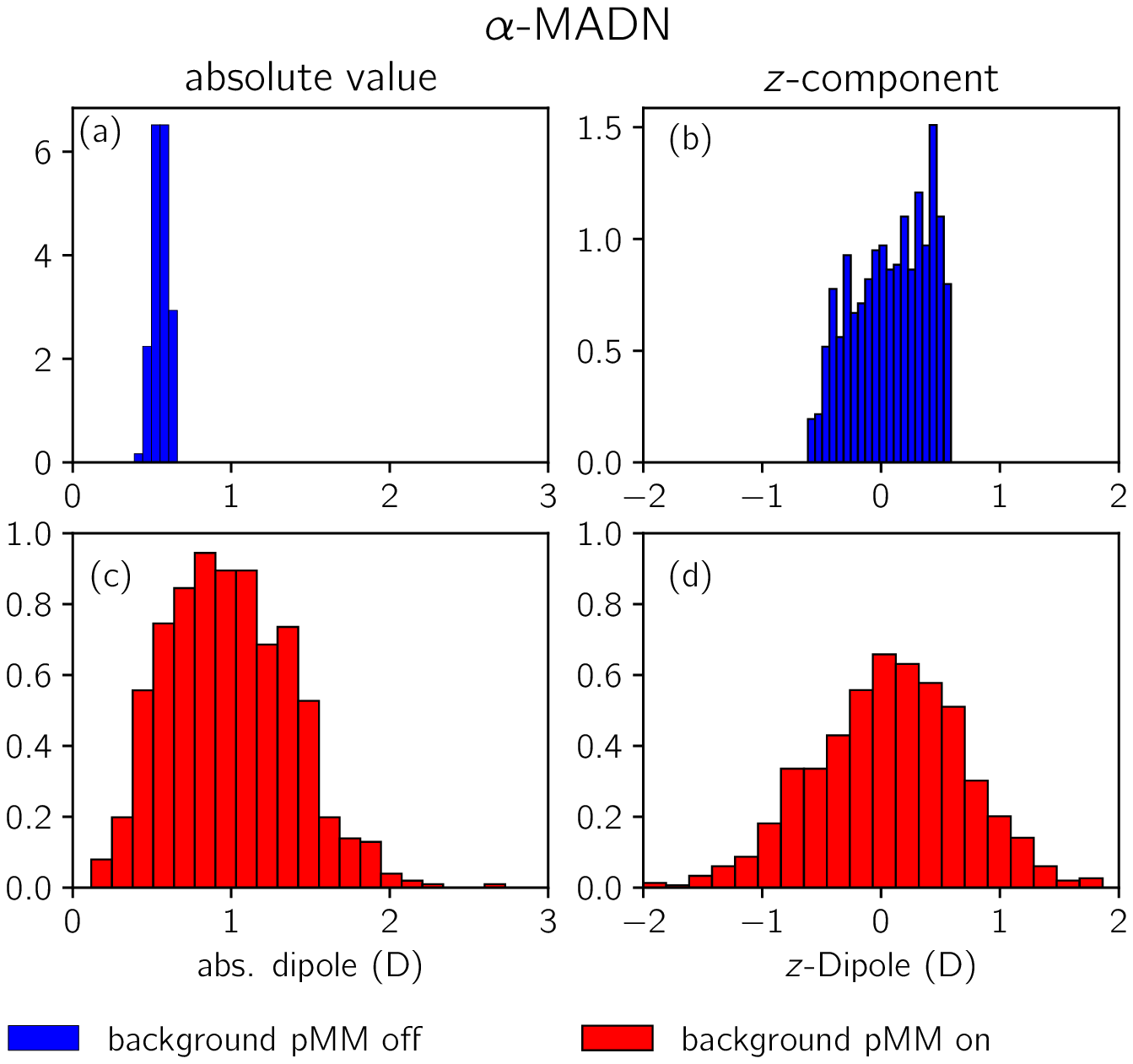}\hfill
    \includegraphics[width=0.49\linewidth]{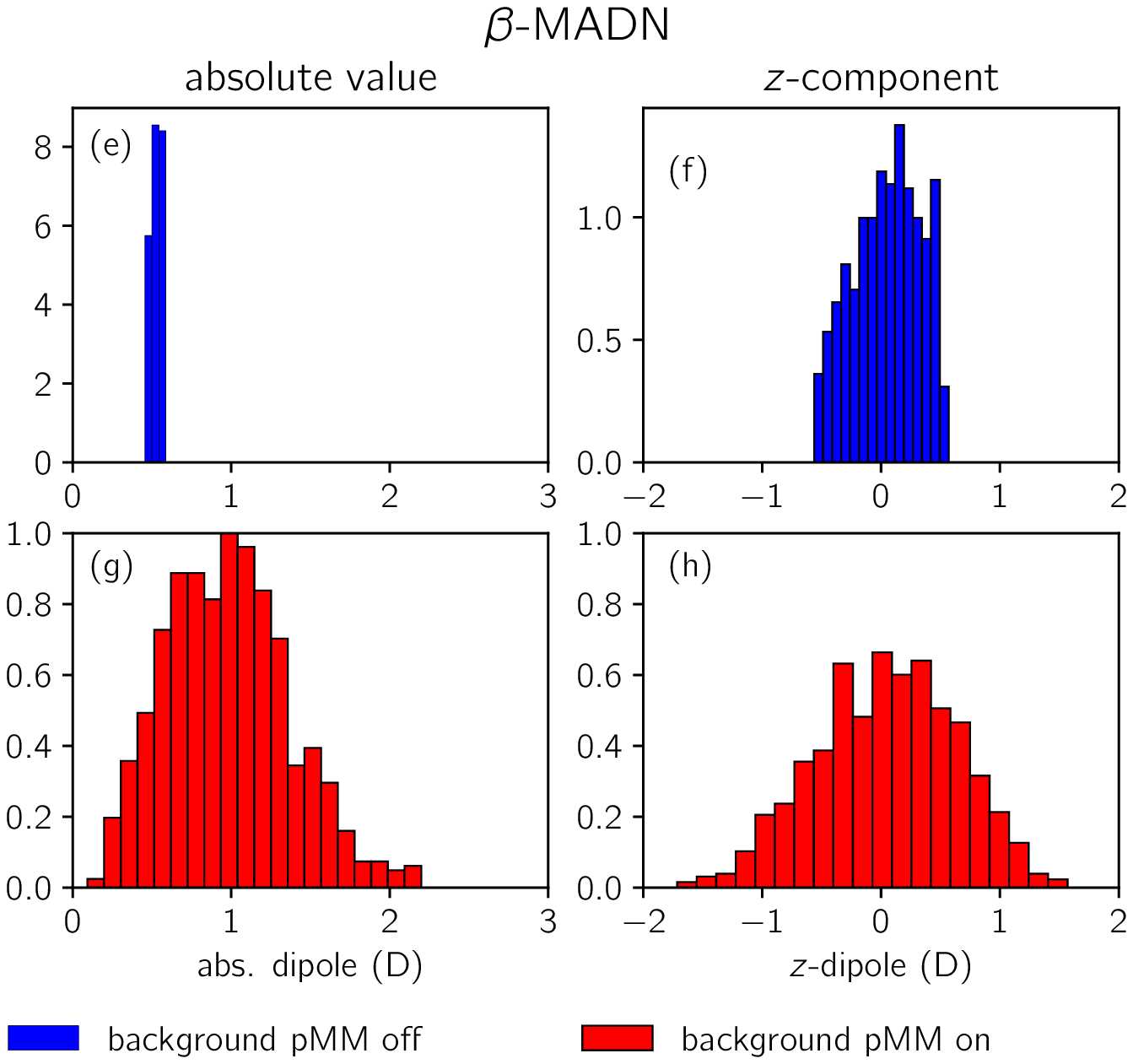}
    \caption{Distributions of the absolute value and $z$-component of molecular dipoles for both \alphamadn and \betamadn inside the amorphous films. Without the pre-polarization of the periodic neutral background (background pMM off, (a) and (b) for \alphamadn and (e) and (f) for \betamadn) both distributions are quite narrow. When background pMM is on ((c) and (d) for \alphamadn, (g) and (h) for \betamadn)  induction effects tend to smear out the dipoles' orientation and strength. See text for further discussion.}
\label{fig:suppinfo:dipoles}
\end{figure}

\begin{figure}[tb]
\centering
  \includegraphics[width=0.49\linewidth]{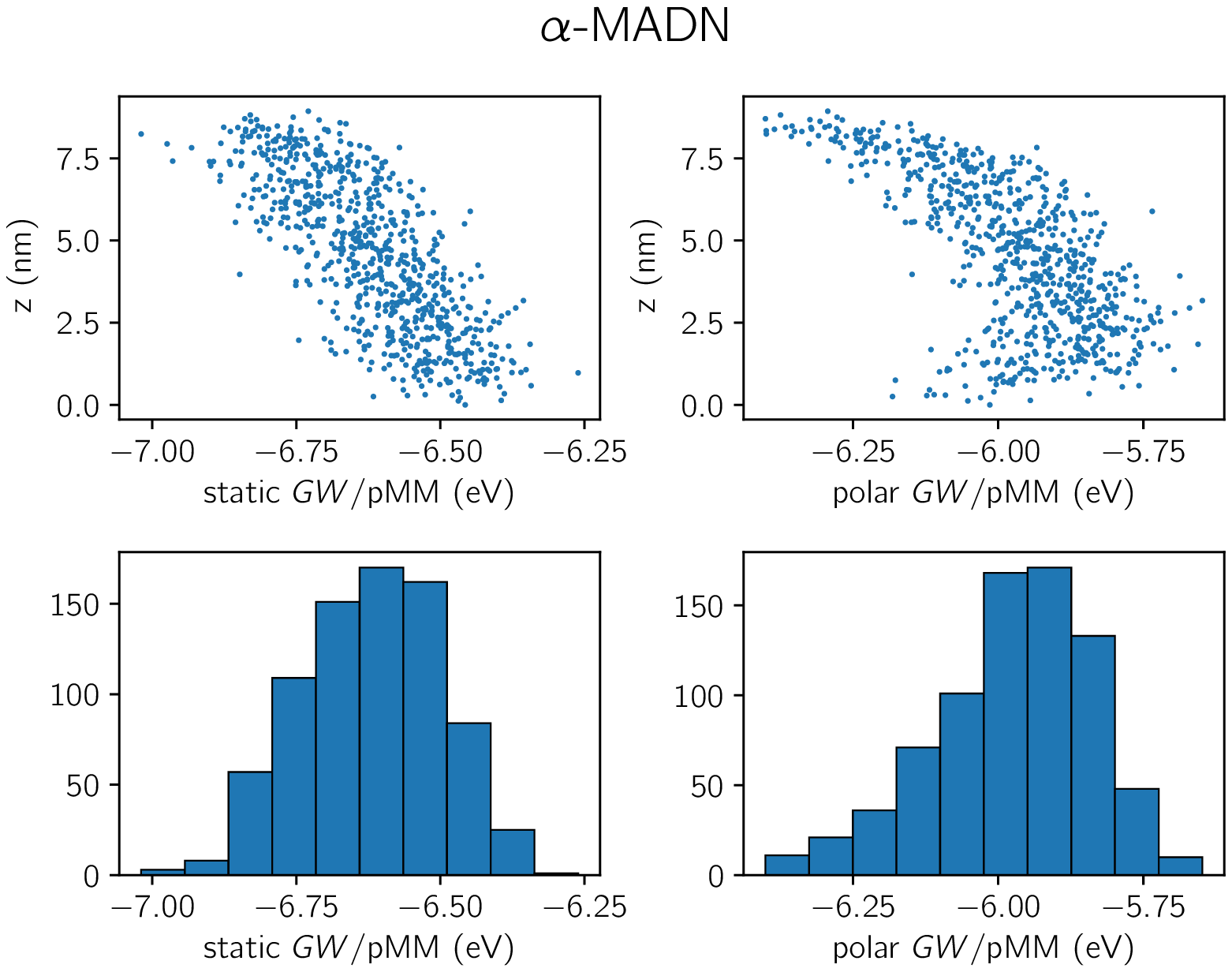}\hfill
 \includegraphics[width=0.49\linewidth]{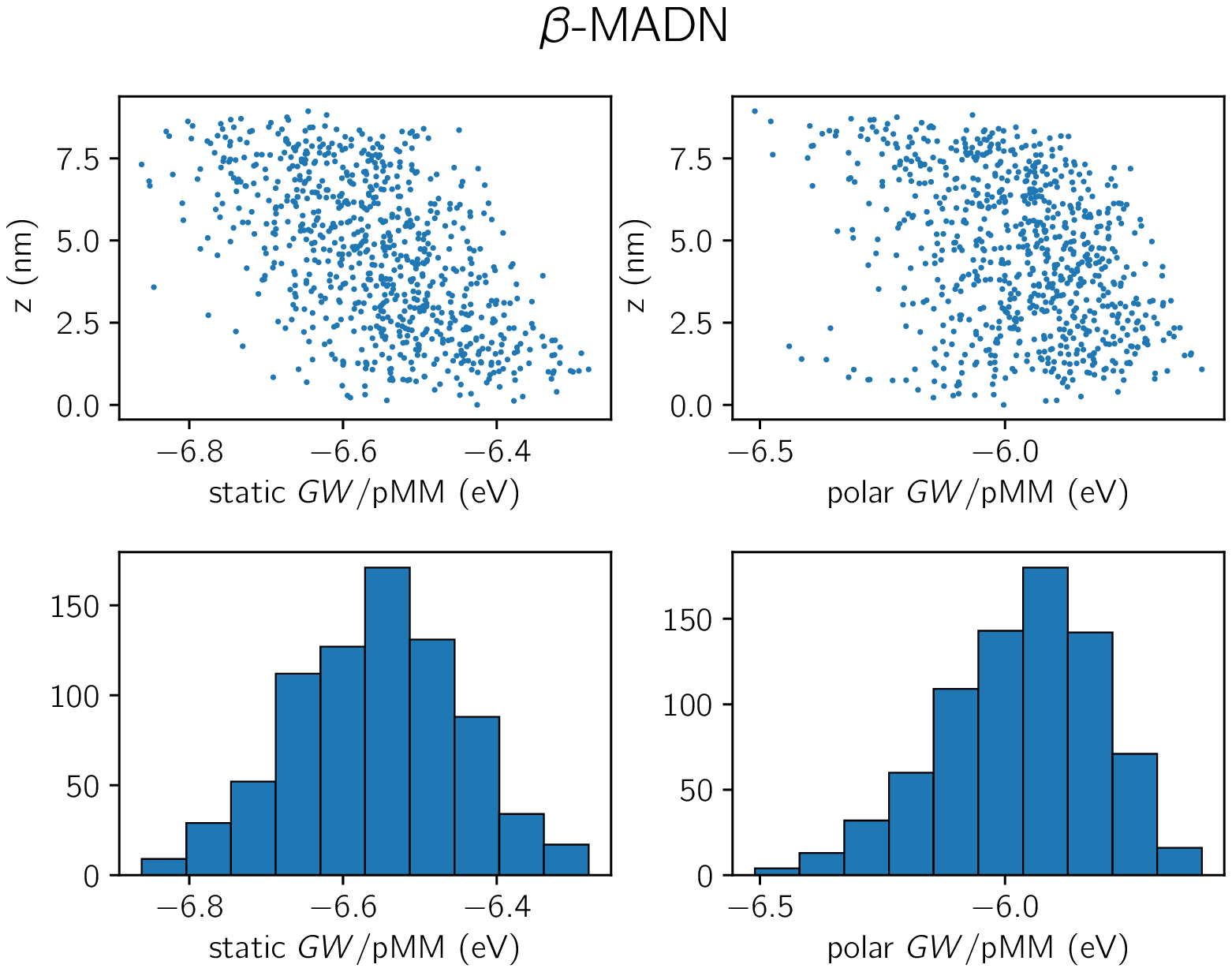}
    \caption{Energies of the HOMO as obtained by static and polar \gwmm simulations for the thin films of $\alpha$-MADN and $\beta$-MADN, respectively. The upper panels show the energies resolved according to the $z$-component of the individual molecule's center-of-mass, while the lower panels show the total energy distribution (or the total density-of-states) in the respective films.}
    \label{fig:suppinfo:embeddingcorrelations}
\end{figure}

In spite of the small effective reduction of the accumulated thin film dipole moment, the distributions in~\fig{suppinfo:dipoles} suggest that the electrostatic potential inside the film and its surfaces shows strong local variations. These manifest themselves directly in a substantial amount of disorder in the calculated excitation energies within the \gwmm approach.
Layer-averaged profiles of the obtained HOMO energies are shown in Fig.~4 of the main text. \Fig{suppinfo:embeddingcorrelations} shows the individual HOMO energies as a function of the $z$-coordinate of the molecule's center-of-mass, as well as their total distributions.

\section{Estimation of the Electron Inelastic Mean Free Path}
\label{sec:imfp}
As mentioned in the main text, we consider the inelastic mean free path (IMFP) of the electrons, $\lambda_\text{in}$, as an upper limit to the electron attenuation length. In the following we focus on estimates of the IMFP within the modeling framework established in this work. Throughout this section, atomic (Hartree) units are used ($\hbar$ = 1, $m_\textrm{e}$ = 1, and $e^2/(4 \pi \epsilon_0) = 1$, with $m_\textrm{e}$ the electron mass, $e$ the elementary charge, and $\epsilon_0$ the vacuum permittivity).

The IMFP is the mean distance between successive inelastic collisions experienced by an electron in a material. Its energy dependence can be estimated with the help of the Energy Loss Function (ELF),
\begin{equation}\label{equ:suppinfo:elf}
  Y_{\text{ELF}}(q,\omega) \equiv \operatorname{Im} \bigg[ \frac{-1}{\epsilon(q,\omega)} \bigg] = \frac{\epsilon_{2}(q,\omega)}{\epsilon_{1}(q,\omega)^{2} + \epsilon_{2}(q,\omega)^{2}},
\end{equation}
where $\epsilon_1(q,\omega)$ and $\epsilon_2(q,\omega)$ are the real
and the imaginary parts of the dielectric function, respectively. The ELF represents the probability of a material to absorb energy $\hbar \omega$ and momentum $\hbar q$ from an energetic incoming particle, such as a photon or an electron with kinetic energy $E_
\text{k}$. The IMFP is related to the ELF via
\begin{equation}
  \lambda^{-1}_\text{in}(E_\text{k}) = \frac{1}{\pi E_\text{k}}
  \int_{\omega_\text{min}}^{\omega_\text{max}} \int_{q_{-}}^{q_{+}} \frac{1}{q} \, Y_{\text{ELF}}(q,\omega) \,dq \, d\omega,
  \label{equ:IMFP}
\end{equation}

where $\omega_\text{min}=E_\text{gap}$, $\omega_\text{max}=(E_\text{k} +E_\text{gap})/2$, and $q_{\pm} = \sqrt{2E_\text{k}}\pm\sqrt{2(E_\text{k}-\omega)}$.

The first step is to compute the ELF. To this end we firstly evaluate
$\epsilon_2(0,\omega)$ according to the non-interacting electron-hole
picture in the Random-Phase Approximation (RPA) as
\begin{equation}
	\epsilon_2(0,\omega) = 16 \, \pi^2 \, \sum_{v,c} \vert \bra{\phi^{\qp}_v}\hat{D}\ket{\phi^{\qp}_c} \vert^2  \delta(\omega-\varepsilon_c+\varepsilon_v), 
	\label{equ:RPA}
\end{equation}
where the sum runs over the occupied ($v$) and unoccupied states ($c$) and
$\hat{D}$ is the dipole moment operator. The real part of the full dielectric function is then obtained using the
Kramers-Kronig relation. From this one can straightforwardly obtain the ELF in
the optical limit ($q \to 0$) using \equ{suppinfo:elf}.

Extending the ELF into the finite-$q$ region is achieved using a model
in which the dielectric response of the system is given by a summation of
non-interacting component oscillators. In the RPA, valence electrons
in the material are approximated by a non-interacting homogeneous gas
where the plasmon energy is expanded to the second order in $q$
\begin{equation}
	Y_{\text{ELF,DL}}(q,\omega,\omega_\text{p}) = \frac{\gamma \, \omega_\text{p} \omega}{(\omega^2 -(\omega_\text{p}+\omega(q))^{2})^2 + (\gamma \, \omega)^2},
	\label{equ:suppinfo:drudelorentz}
\end{equation}
where $\omega(q) = E_\text{gap} + \alpha q^2$ and $\gamma$ is the damping coefficient.
The above optical Drude-Lorentz (DL) ELF in the form of
\equ{suppinfo:drudelorentz} has a singularity at the plasma frequency $\omega_\text{p}$.

\begin{figure}[tb]
\centering
   \includegraphics[width=0.49\linewidth]{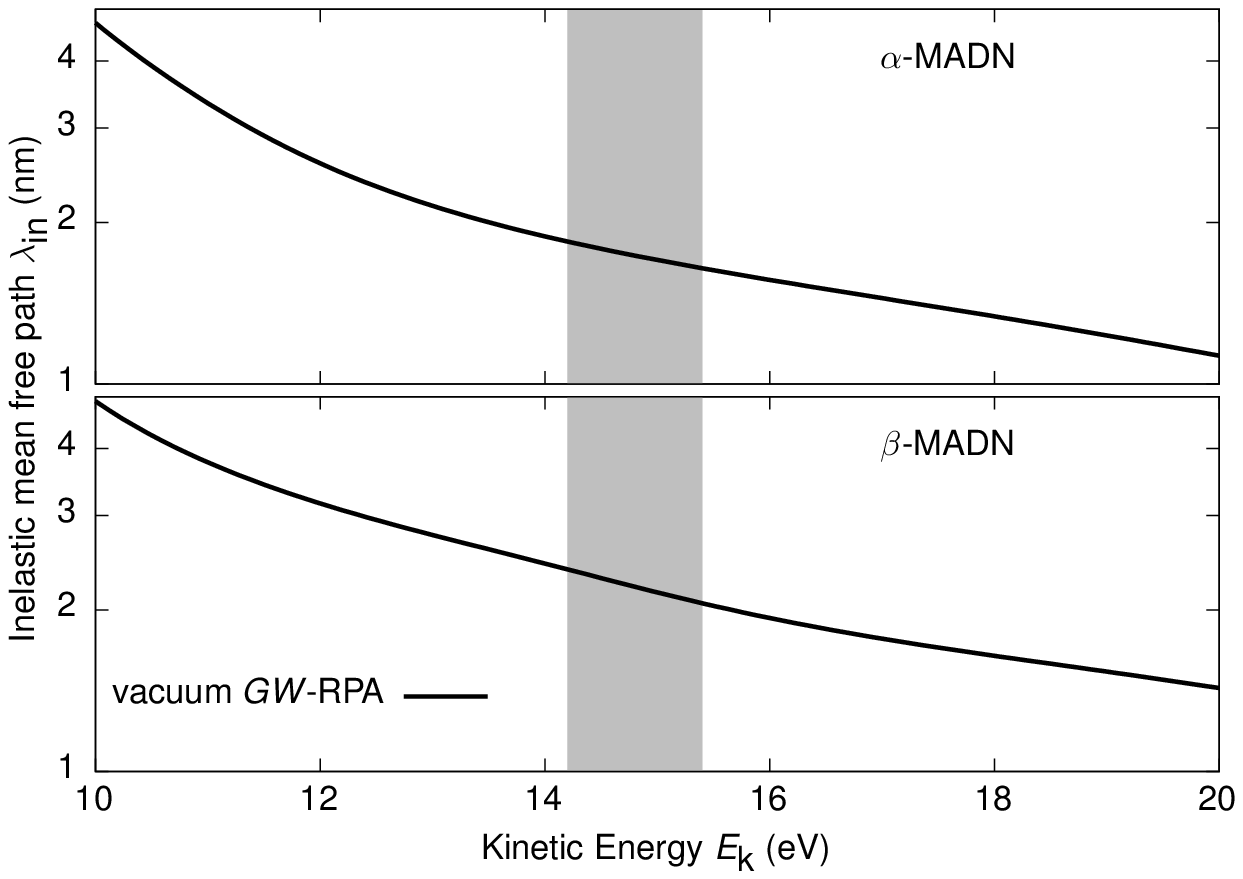}
    \caption{Kinetic energy dependence of the inelastic mean free path in \alphamadn and \betamadn, obtained using the RPA with vacuum $GW$ quasiparticle energies. The gray shaded area indicates the range of interest (\unit[14.2]{eV} - \unit[15.4]{eV}).}
    \label{fig:suppinfo:imfp}
\end{figure}

To extend this approach from a non-interacting to an interacting medium, we
consider the optical ELF as composed of DL-ELF terms with closely-spaced plasma frequencies $\omega_i$ such that
\begin{equation}
	Y_{\text{ELF}}(0,\omega) = \sum_{i} A_{i} \, Y_{\text{ELF,DL}}(0,\omega,\omega_\text{p}=\omega_i) .
\end{equation}
Once we have found the amplitude parameters $A_i$ via a fitting procedure to
our calculated ELF, we can build a momentum-dependent ELF according to
\begin{equation}
        Y_{\text{ELF}}(q,\omega) = \sum_{i} A_{i} \, Y_{\text{ELF,DL}}(q,\omega,\omega_\text{p}=\omega_i),
        \label{equ:suppinfo:qELF}
\end{equation}
with the extension to finite $q$ as in~\equ{suppinfo:drudelorentz}. Entering~\equ{suppinfo:qELF} into~\equ{IMFP}, we perform the integration over $q$ and $\omega$ numerically to obtain $\lambda_\text{in}(E_\text{k})$. For He-I UPS (photon energy \unit[21.2]{eV}) and
with $\varepsilon_i$ in the range of \unit[-7.0]{} to \unit[-5.8]{eV} (see Fig.~4 of the main text), the kinetic
energy of interest is approximately 14.2~$-$~15.4~eV. \Fig{suppinfo:imfp} shows the kinetic energy dependence of the IMFP as obtained with vacuum quasiparticle energies. For $E_\text{k}=\unit[15.0]{eV}$ we obtain IMFPs of \unit[1.69]{nm} for \alphamadn and \unit[2.14]{nm} for \betamadn.

These values should be considered as upper limits to the real IMFP, and hence also the EAL, due to the neglect of, e.g., changes in the full quasiparticle spectrum due to morphology effects, intermolecular excitations in the RPA, or excitonic effects. Furthermore, elastic processes can additionally reduce the electrons' mean free path and hence the attenuation length. Explicit inclusion of these additional scattering mechanisms in our estimates is beyond the scope of this work.

From coverage-dependent studies of the He-I (\unit[21.2]{eV}) UPS spectrum of MADN on Au, we find (1) that the Au-contribution to the spectrum has not yet decreased significantly for a coverage of \unit[0.6]{nm} ($\alpha$-MADN), (2) that this contribution has decreased to about 10\% for a coverage of \unit[1.2]{nm} ($\beta$-MADN), and (3) that this contribution has almost vanished for a coverage above \unit[1.6]{nm} ($\alpha$-MADN)~\cite{christ_2019}. Given the uncertainties in obtaining the attenuation length from theory, detailed above, and based on these experimental observations, we regard the value of the attenuation of \unit[1]{nm}, adopted in the main text for both isomers, as a fair estimate.

